# VOC Sensing Using Batch-fabricated Temperature Compensated Self-Leveling Microstructures

R. Likhite[a], A. Banerjee[a], A. Majumder[a], M. Karkhanis[a], H. Kim[a], and C.H. Mastrangelo[a,*]

[a] *Department of Electrical and Computer Engineering, University of Utah, Salt Lake City, USA*

**Abstract**

We present the design, fabrication, and response of a low-power, polymer-based VOC sensor based on the self-leveling of mechanically leveraged structures. The device utilizes folded polymer-coated microcantilevers to achieve passive temperature compensation without the need for additional compensating sensors or electronics. We demonstrate that a self-leveling vapor sensor provides the same gas response as a simple microcantilever geometry, showing ~20% change in device capacitance when subjected to 35-85 %RH change while showing nearly-zero baseline drift due to changes in ambient temperature when the temperature is increased from 23-72°C which is ~52-fold better than a simple microcantilever geometry. The response of the VOC sensor was measured using three polymers (Polyimide, Polyurethane, and PDMS) against five different analytes (Ethanol, Acetone, Benzene, Hexane, and Water) and an SVM-based model was used to show target specificity. The sensor also showed an absorption response time ($\tau_{90}$) of ~138s. We propose that the self-leveling vapor sensor geometry is a significant improvement to a simple microcantilever vapor sensor as it offers the same performance but shows near-complete elimination of temperature-induced baseline drift.

## 1. Introduction

In recent years, deployment of IoT (Internet of Things) nodes for various applications [1–3] has seen enormous growth, which has fueled the demand for low-power sensors at the back-end [4–9]. One such application is IAQ (Indoor Air Quality) monitoring for the detection of VOCs (Volatile Organic Compounds). VOCs are chemicals that have a very high vapor pressure at room temperature. Therefore, these compounds are usually in their vapor phase in the environment, which increases their chances of getting inhaled by humans and causing acute and chronic health issues [10,11].

VOC sensing for indoor air quality monitoring is done using various types of chemical sensors and electronic nose systems [12–15]. VOC sensing using polymer-coated microcantilevers beams has also been demonstrated previously [16–20], but these devices typically have a high limit of detection and, are cross-sensitive to humidity and temperature compared to the other types of sensors. However, they do offer significant advantages for low-cost, low-power, and portable applications as demanded by IoT-based frameworks. For example, microcantilever-based sensors have a small size, show linear behavior, are easy to batch fabricate, and, most importantly, consume nearly zero DC power when used with CMOS integrated capacitive readouts.

Techniques for improving the limits of detection of polymer-based microcantilevers type vapor sensors operating in static-mode have been demonstrated earlier [21–24] but, the issue of baseline drift due to changes in ambient temperature still plagues these sensors. These devices are inherently bi-material in nature, which makes them extremely sensitive to output fluctuations due to small changes in their operating temperature. Conventional solutions for compensating temperature-induced baseline drift include, use of a differential arrangement [25,26], active temperature stabilization systems [27,28] or software compensation based on pre-determined temperature response characterization [29]. All of these require power for the compensating electronics which limits their use in applications where power is not available. VOC sensor systems for ultra-low-power applications can be developed if MEMS-based [30–33] temperature compensation methods are used instead.

In this paper, we report for the first time, the design, fabrication, and testing of a new type of low power, batch-fabricated VOC sensor with improved temperature compensation based on mechanical self-leveling [30]. This article expands on a proof of concept shown earlier [34] and presents extensive characterization of the sensor response to analytes using different sensing polymers.

*Corresponding Author – Carlos H. Mastrangelo
 Email – carlos.mastrangelo@utah.edu

## 2. Baseline drift in microcantilever sensors

Fig. 1a shows a schematic of a typical microcantilever type vapor sensor based on polymer swelling. This device consists of a gas-insensitive microcantilever coated on top with a gas-sensitive polymer. When such a device is exposed to an analyte vapor such as VOCs or water vapor, the sensing polymer absorbs the analyte molecules and swells. Since the polymer is mechanically attached to the microcantilever, the swelling generates a surface stress [35] and hence, a bending moment $M$. This produces uniform bending, which translates into a downward deflection of the microcantilever tip. The tip deflection can be detected using various transduction techniques [18,36–38]. For example, if a fixed electrode is placed underneath the tip of the bending microcantilever, beam bending can be sensed using a capacitive readout as the top electrode (beam) moves freely over a fixed bottom electrode [37] as shown in Fig. 1b.

For a microcantilever beam of length $L$ and, thickness $t_c$ coated with a sensing polymer of thickness $t_p$, the deflection of the beam tip $y$, due to any misfit strain $\varepsilon_m$ between the two layers can be given by [39],

$$y = \frac{1}{2}\kappa . L^2 \tag{1}$$

Where, $\kappa$ is the beam curvature produced by the moment $M$, and expressed as,

$$\kappa = \frac{6 E_p E_c (t_c+t_p) t_c t_p}{E_p^2 t_p^4 + 4 E_p E_c t_p^3 t_c + 6 E_p E_c t_p^2 t_c^2 + 4 E_p E_c t_p t_c^3 + E_c^2 t_c^4} \cdot \varepsilon_m \tag{2}$$

Where $E_p$ and $E_c$ are the Young's Modulus of the polymer and the microcantilever, respectively. The misfit strain $\varepsilon_m$ can be produced by both, a change in vapor concentration and a change in ambient temperature such that,

$$\varepsilon_m = (\alpha_c - \alpha_p).\Delta T + \beta_p . C_G \tag{3}$$

Where, $\alpha_p$ and $\alpha_c$ are the CTE (Coefficients of Thermal Expansion) of the polymer and microcantilever material respectively, $\Delta T$ is the change in ambient temperature, $\beta_p$ is the vapor-sensitive expansion coefficient of the sensing polymer and, $C_G$ is the vapor concentration. Since polymers used for vapor sensing typically have a high CTE [40–42], which is many folds higher than that of the structural material of the microcantilever, which is usually silicon [43], the microcantilever vapor sensor is a bimaterial stack which can also undergo uniform bending, due to changes in ambient temperature. As the temperature increases in the absence of an analyte vapor, the difference in CTE of the two materials produces a misfit strain in the device and tip deflection, which is visible as baseline drift in the output. Simple microcantilever structures cannot therefore decouple the effects of stimulus from temperature and analyte vapor.

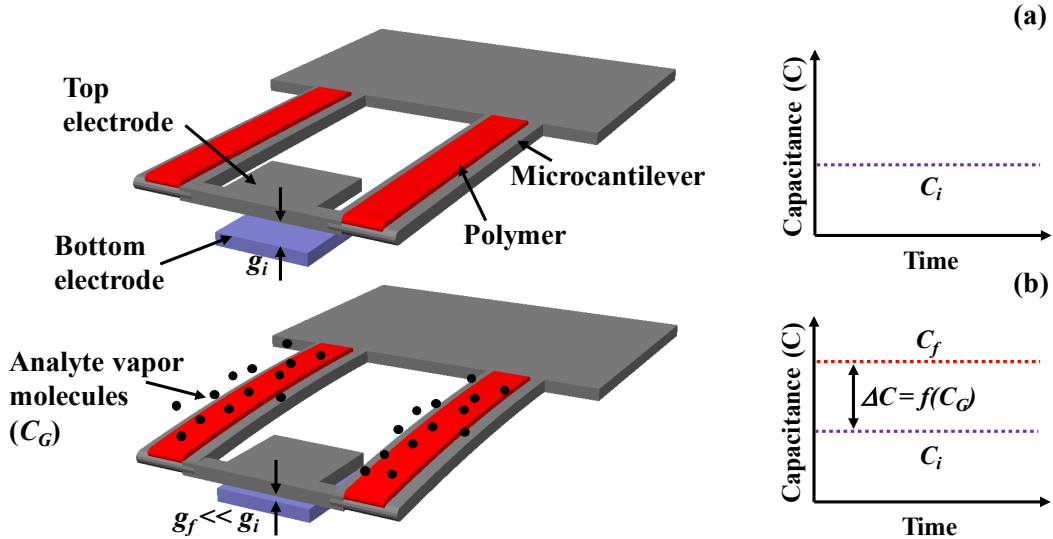

**Figure 1:** Operation of a reference polymer-based capacitive microcantilever vapor sensor (a) Before exposure to analyte vapors. (b) After exposure of the device to analyte vapors concentration, $C_G$. The sensor is composed of two microcantilever beams to keep its spring constant same as the self-leveling design, presented in the next section.

## 3. Self-leveling VOC sensor

In this work, we have developed a microcantilever type vapor sensor which can achieve temperature compensation by operating in two separate mechanical deflection modes. The sensor has a common mode response and a differential mode response. The common mode response is temperature dependent, but the differential mode response is made gas dependent. The mechanical structure is configured such that the overall tip deflection rejects the common mode temperature response but is sensitive to the differential response. The common mode temperature response is therefore rejected, and the differential mode vapor response is accepted, leading to successful decoupling of the terms in Eq$^n$. (3). This is achieved using a self-leveling micromechanical vapor sensor, as shown in Fig. 2. Self-leveling geometries have been previously utilized to implement temperature compensated infrared detection sensors [30,31]. However, this is the first report of a VOC sensor based on a self-leveling geometry.

A self-leveling vapor sensor is implemented using two sets of U-shaped, folded microcantilever beams coated with a sensing polymer. The sensing polymer is capable of absorbing the analyte vapor and swelling similar to that in simple microcantilever vapor sensor arrangement described earlier. The polymers on the beams are covered with an ultra-thin $Al_2O_3$ layer deposited using a PA-ALD (Plasma-Assisted Atomic Layer Deposition) process, which acts as a gas diffusion barrier [44]. The diffusion barrier on the polymers on the inner beams is patterned to provide access holes to expose part of the polymer to the external environment, as shown in Fig. 2. The inner beams are connected to a suspended top sense electrode, which floats over a fixed bottom electrode near the anchor thus, forming a variable capacitor for output transduction.

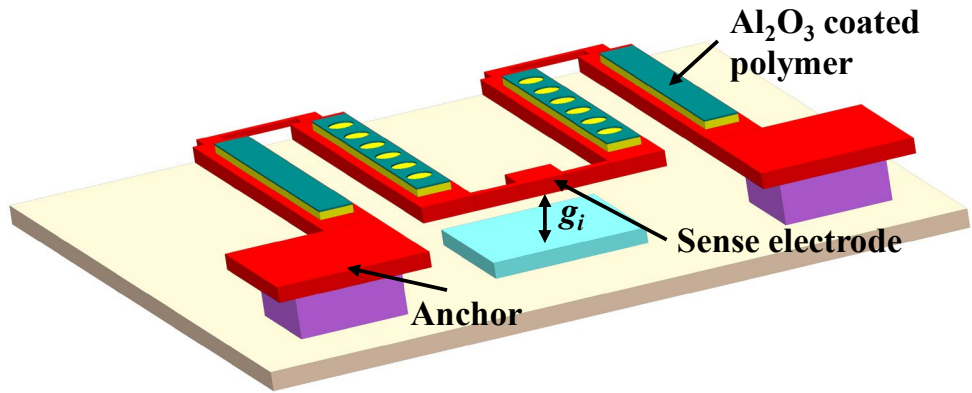

**Figure 2:** Schematic of the self-leveling vapor sensor

### 3.1 Working principle

Fig. 3(a-d) shows the response of a self-leveling sensor to a change in ambient temperature and analyte vapor concentration, respectively. When the sensor is exposed to a fluctuation in ambient temperature, all the four polymer-coated microcantilever beams act as bimetallic strips and bend down, as shown in Fig. 3a. This brings the free ends of the inner beams back to its original position near the anchor leading to a negligible change in gap w.r.t the fixed bottom electrode, and hence, the sensor output is constant as shown in Fig. 3b. When the gas/vapor concentration around the device changes, the diffusion barrier on the polymers produces a differential response from the outer and inner beams. Since the polymers on the outer beams do not swell, the gas-insensitive outer microcantilever beams do not bend and, the free end of these beams acts as a virtual anchor for the inner beams. The presence of access holes in the diffusion barrier on the inner beams produces a polymer swelling effect in response to a gas/vapor and results in a downward deflection of the top sense electrode due to virtual anchoring of the folded end of the beam as depicted in Fig. 3c. This can be sensed as a change in capacitance of the device as shown in Fig. 3d. The self-leveling vapor sensor is, therefore, able to successfully decouple the effects of temperature and gas/vapor on its output thus, producing a temperature compensated vapor sensor with zero external power. Note that, the mechanical properties of the composite beams (poly-Si, polymer and $Al_2O_3$) can be affected if the thickness of the gas diffusion barrier is made very high as the three layers have significantly different Young's Modulus. However, in the self-leveling VOC sensor, the effect of the $Al_2O_3$ layer on the beam bending is minimal as the thickness of the diffusion barrier is extremely small compared to the thickness of the poly-Si and the sensing polymer.

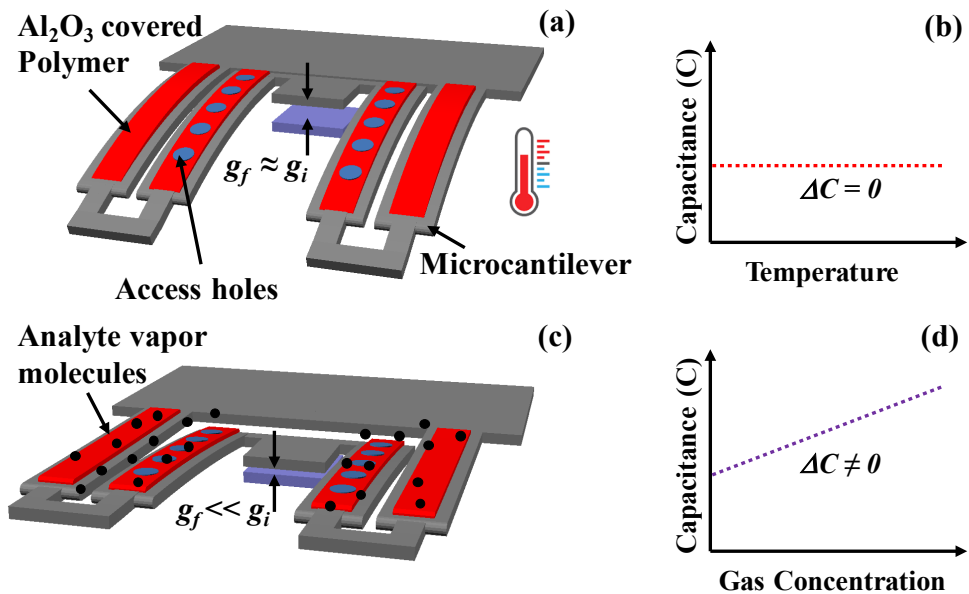

**Figure 3:** Dual-mode response of the self-leveling sensor. (a) Device response to temperature (b) Rejection of common-mode response (c) Device response to gas (d) Acceptance of differential mode response.

*3.2 Target Selectivity*

Since polymers typically absorb a wide range of analytes, utilization of a single polymer-based device is not suitable for selective detection of a target VOC. A suite of sensors with different sensing polymers operated in conjunction with each other can be used to construct a 'fingerprint' vector response to different analytes. This concept is widely used in developing an electronic nose system [45,46]. The response of multiple sensors to a particular analyte is obtained, and clustering algorithms [47,48] are implemented to create a unique signature that can identify a chemical. Therefore, e-nose systems based on low-power vapor sensors are becoming popular candidates in applications such as food quality monitoring [49], odor sensing [50], and air-quality monitoring [51,52].

In this work, we have microfabricated self-leveling VOC sensors using three sensing polymers, i.e., Polyimide, Polyurethane, and PDMS. The devices were tested against various analytes vapors and the gas targets were next identified and quantified using a support vector machine (SVM) machine learning technique [53,54].

**4. Fabrication and Imaging**

*4.1 Device Fabrication*

Fig. 4(a-l) shows a simplified fabrication procedure of the device. The process starts by wet thermal oxidation of a 4" silicon wafer. A thin layer of 200nm low-stress $Si_3N_4$ is then deposited over the wafer using LPCVD (Low-Pressure Chemical Vapor Deposition) process (Fig. 4a). The nitride and silicon dioxide layers are used to isolate the sensors from the substrate. This bi-layer stack is then patterned using conventional UV photolithography with S1813 photoresist, and a RIE (Reactive Ion Etching) etch process using $CF_4/O_2$ plasma. The photoresist is then stripped off, the wafer is thermally oxidized, and another layer of $Si_3N_4$ is deposited. This process forms an isolation trench for the bottom electrode, as shown in Fig. 4b. The bottom fixed electrode for the device is then formed by depositing poly-Si using LPCVD process and doped using a $P_2O_5$ solid-source diffusion process. The poly-Si layer is patterned and etched using $SF_6$ RIE, as shown in Fig. 4c. A layer of 4 μm thick PSG (Phosphosilicate Glass) is then deposited over the wafer using LPCVD (Fig. 4d) to serve as a sacrificial layer for the devices. Dimples are patterned using photolithography and BOE (Buffered Oxide Etch) wet etching (Fig. 4e) to prevent device failure due to stiction [55]. The PSG is also patterned using $CF_4/O_2$ RIE etch process to create anchors for the MEMS devices. Note that, in case of the self-leveling devices, the thickness of the sacrificial layer was reduced to 1 μm by using an RIE process to increase the initial capacitance of the device. The structural layer for the devices is formed by depositing 6 μm thick

poly-Si using LPCVD and doped using a diffusion process as shown in Fig. 4f. The folded links of the self-leveling device were first pattered using photolithography, and the poly-Si was partially etched using $SF_6$ RIE to reduce the thickness to 4 µm. This was done to make the folded link stiffer than the microcantilever beams, thus preventing beam twisting. The next step is deposition of the sensing polymer using a spin-coating process. For this work, three wafers were processed simultaneously, and three sensing polymers, i.e. Polyimide, PU (Polyurethane) and PDMS (Polydimethylsiloxane) were spin-coated on them. The polyimide used in this work was HD-4104 [40], which is a negative-tone photo definable polyimide. The substrate was first dehydrated by heating on a hotplate at 200°C for 20 mins. The native oxide was then removed using a 10s dip in a BOE solution. A commercially available adhesion promoter, VM-651 was then applied to the wafer using the manufacturer's recommended procedure [56]. The polyimide was then dispensed and spin-coated over the wafer at 3000 rpm for 45 sec. The film was soft-baked on a hotplate at 90°C for 2 mins and 100°C for 2 mins. The polyimide was cured completely in an oven at 300°C for 3 hrs under a $N_2$ environment. The polyurethane used in this work was Bayhydrol-110 [57] which is an anionic dispersion of an aliphatic polyester urethane resin in water/n-methyl-2-pyrrolidone. The substrate preparation steps included wafer dehydration and a native oxide etch similar to the polyimide sample. The Bayhydrol solution was then dispensed on the wafer and spin coated at 4500rpm for 30s. The wafer was placed overnight in an oven at 60°C for a soft-bake. After the overnight bake, final curing of the film was done by placing the wafer on a hotplate at 150°C for 10 mins. For the third sample, Sylgard-184 PDMS was used [58]. The same substrate preparation steps were followed for this wafer as the polyurethane sample to promote film adhesion. A 1:10 ratio mixture of PDMS:Curing agent was first prepared and then degassed in a vacuum desiccator to remove all air bubbles from the solution. The mixture was then diluted with Hexane in a 1:1 ratio by weight and mixed thoroughly. This diluted solution of PDMS was then spin-coated over the sample wafer at 6000rpm for 150s. The film was then cured on a hotplate at 120°C for 20 mins. The PDMS film was further baked at 240°C for at least 24 hrs in an evacuated vacuum oven. This was done to remove residual oil from the film and minimize contamination of cleanroom equipment and vacuum pumps. Table 1 summarizes the process parameters used to spin coat the polymers. The obtained thickness of the sensing polymer was measured using a Tencor P-20h profilometer. After polymer deposition, polyimide and polyurethane devices were patterned (Fig. 4g) using a 200nm thick Al metal hard mask and $O_2$ plasma dry etch. The PDMS coated devices required a $CF_4/O_2$ gas mixture for dry etching. It was observed that the Polyurethane was incompatible with different organic solvents such as Acetone, IPA and methanol which are commonly used during MEMS fabrication. Therefore, extreme precaution was taken during device fabrication to ensure that the polyurethane polymer wafer was not exposed to liquid solvents after the polymer had been deposited. Resist strip-off after subsequent lithography steps was done using a UV flood exposure and then resist removal using the developer AZ 300 MIF. A ~10nm thick layer of $Al_2O_3$ gas diffusion barrier is then deposited over the sensing polymer using a PA-ALD process at room temperature and patterned using a BOE wet etch to create access holes on the inner polymer-coated beams as shown in Fig. 4h. The access holes allowed ~28% of the polymer to get exposed to the outside environment. The thickness of the deposited ALD film was confirmed using Ellipsometry. Next, the remaining 4 µm thick structural poly-Si layer is then patterned using $SF_6$ dry etch to pattern the set of folded beams (Fig. 4i). Before dicing, a layer of 200nm thick Au with 30nm Cr adhesion layer was deposited over the wafer using DC sputtering and 30 µm thick AZ9260 double-coated photoresist is patterned on top of the sensing polymer to protect it from future wet etch processes as shown in Fig. 4j. The wafer is then diced, and the devices are released (Fig. 4k) using a BOE wet etch for 120 mins. After device release, the devices are rinsed in DI water, and the photoresist/metal layer is stripped away. Finally, the devices are rinsed in DI water and allowed to air dry in a ventilated oven as shown in Fig. 4l.

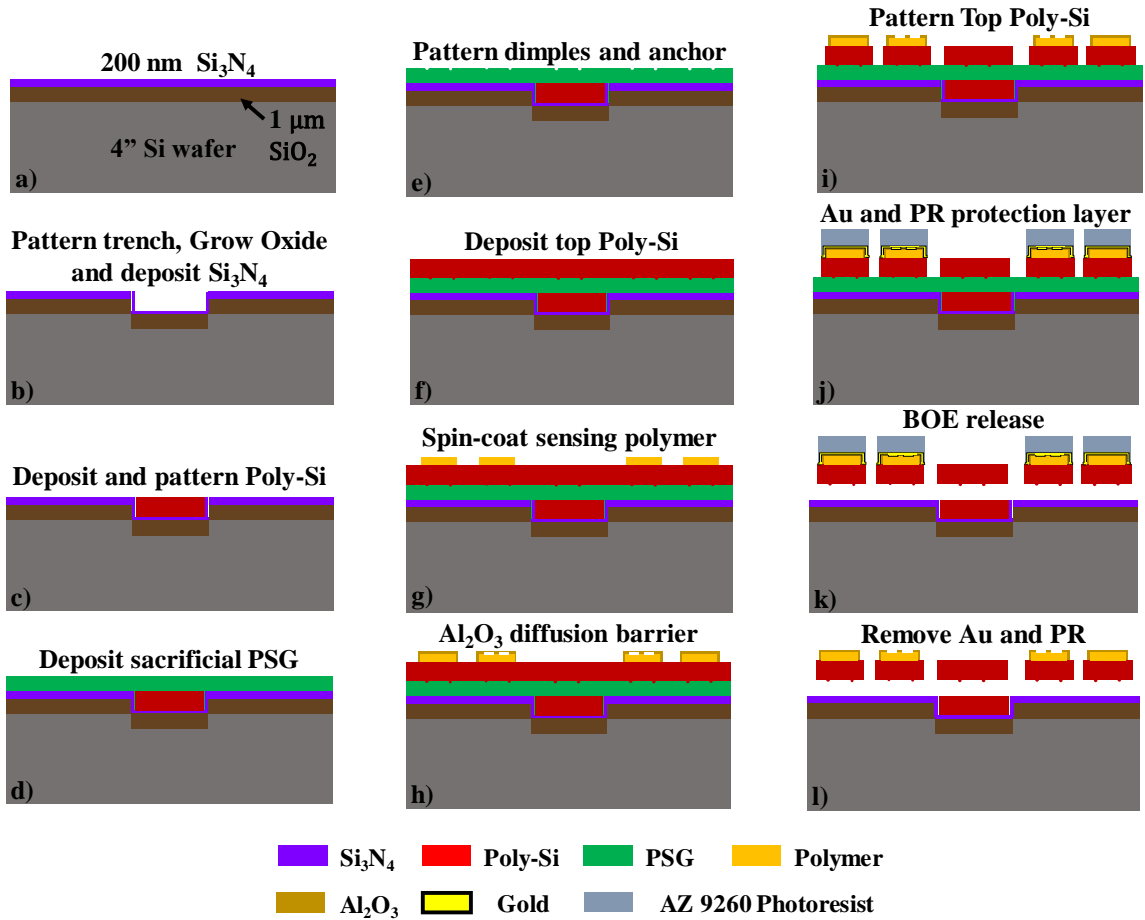

**Figure 4:** Simplified fabrication procedure of the device.

Table 1: Process parameters for different sensing polymers

|  | Polyimide | Polyurethane | PDMS |
|---|---|---|---|
| Chemical | HD 4104 | BAYHYDROL-110 | Sylgard-184 1:10 (PDMS:Curing agent) diluted 1:1 with Hexane |
| Adhesion Promotion | 1) Dehydrate wafer (200°C for 20 mins) 2) Native Oxide etch 3) VM-651 adhesion promoter | 1) Dehydrate wafer (200°C for 20 mins) 2) Native Oxide etch | 1) Dehydrate wafer (200°C for 20mins) 2) Native Oxide etch |
| Spin speed/time | 3000rpm/45s | 4500rpm/30s | 6000rpm/150s |
| 1st Bake | 90°C/120s and 100°C/120s on a hotplate | Overnight in an oven at 60°C | 120°C/20 mins on a hotplate |
| 2nd Bake/Curing | 300°C/3 hrs in $N_2$ oven | 150°C/10 mins on a hotplate | 240°C/24 hrs in a vacuum oven |

| Etch chemistry | O$_2$ Plasma RIE | O$_2$ Plasma RIE | CF$_4$/O$_2$ Plasma RIE |
|---|---|---|---|
| **Obtained thickness** | ~3.35 µm | ~3.10 µm | ~1.9-2.0 µm |

*4.2 Imaging*

High-resolution Scanning Electron Microscope (SEM) imaging of the device was done on an FEI Nova Nano 630 SEM at an accelerating voltage of 10.0 kV to verify the fabricated device structure. Fig. 5 and 6 show SEM images of a self-leveling sensor with PDMS polymer and a reference uncompensated microcantilever beam with PDMS, respectively. Fig. 5a shows the self-leveling sensor with Al$_2$O$_3$ coated polymers. Polymers on beam 1 and 4 are entirely covered with Al$_2$O$_3$, and the diffusion barriers on beam 2 and 3 are patterned to provide access holes. Fig. 5b shows a zoomed-in image of the suspended top plate of the capacitive sensor with some access holes to allow the wet release procedure. The top electrode area was 3025 µm$^2$, and the gap between the top and bottom electrodes was 1 µm. The microcantilever beams were 450 µm long, 4 µm thick and 50 µm wide. The folded links were 6 µm thick. The polymer patches on top of the beams were 30µm wide. Fig. 6a shows the SEM image of the uncompensated reference microcantilever beam of the same dimensions. The polymer on the reference cantilever beam is also covered with a diffusion barrier patterned with access holes for a fair comparison to the self-leveling device. Note that the uncompensated beam consists of four capacitors connected in parallel to provide the same base capacitance as the self-leveling sensor as the air gap in the reference cantilever beams was 4µm, as shown in Fig. 6b. This was done to prevent the beams from touching the substrate due to downward beam bending during a temperature sweep of the device.

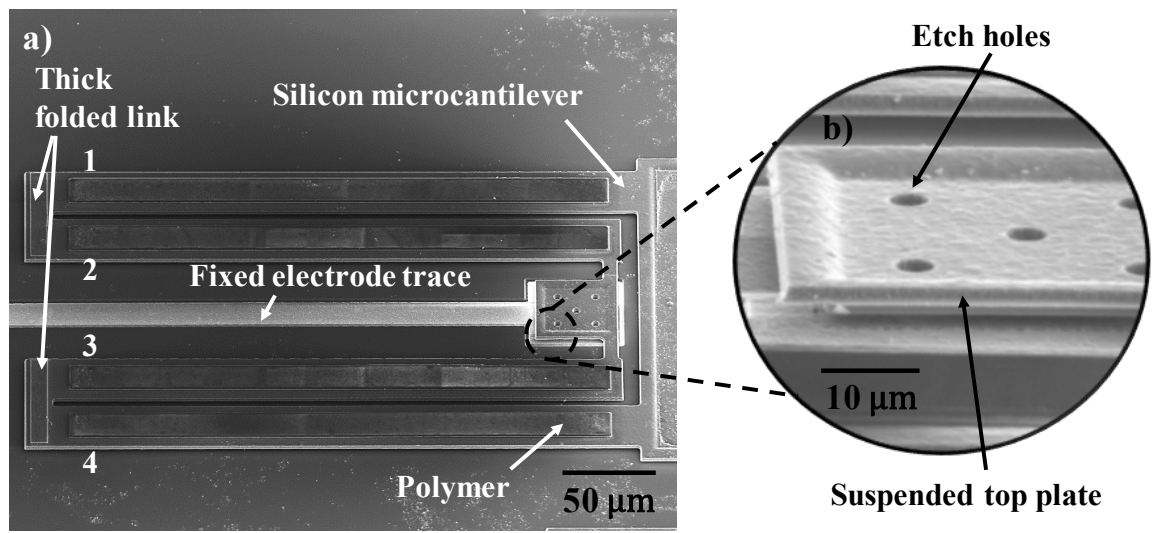

**Figure 5:** SEM image of the Self-leveling sensor a) 450µm long SL device with PDMS polymer b) Zoomed-in image of the suspended top electrode of the device.

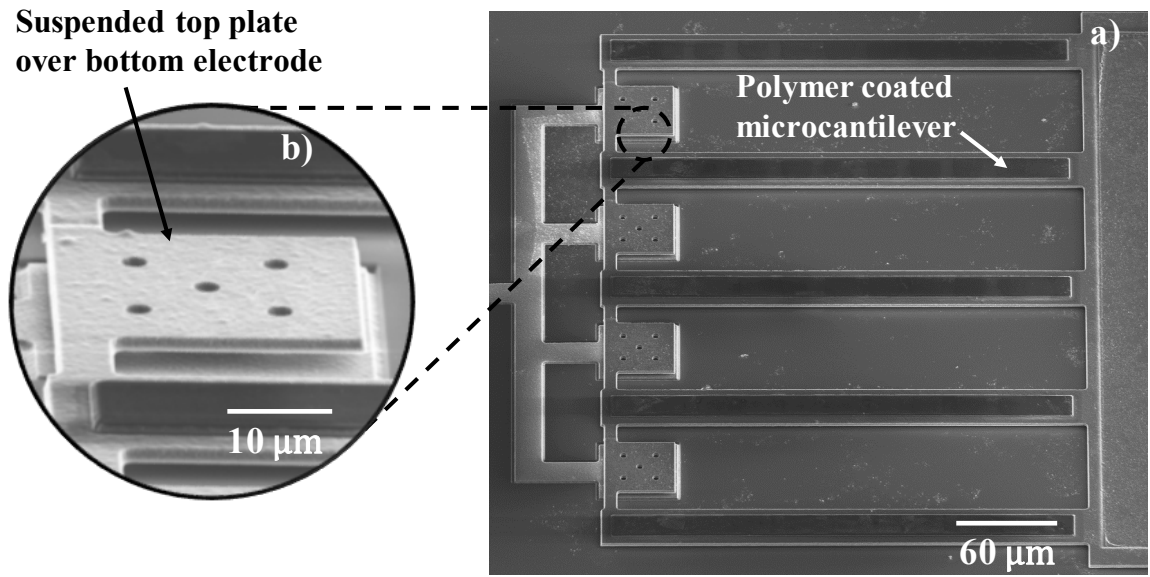

**Figure 6:** SEM image of the reference microcantilever a) 450μm long reference device with PDMS polymer. b) Zoomed-in image of the suspended top electrode of the device.

## 5. Testing and Characterization

*5.1 Test setup and procedure*

The sensor testing was done at a probe station enclosed in a metallic box. The enclosing box was grounded to reduce outside interference and noise during measurement. The device capacitance was measured as a function of changing gas concentration using an HP 4284a LCR meter connected to the probe station at 1 MHz frequency using a 100 mV AC signal with long integration. The layers on the backside of the chip were scratched off, and the substrates were grounded by using silver paint to connect the chips to the grounded chuck of the probe station. The base capacitance of the devices was measured to be ~70 fF. Two types of testing arrangements were used for this experiment, as described below.

*5.1.1 Relative Humidity and Temperature testing*

In order to validate the distinct gas and temperature response of the self-leveling sensor, the device performance was compared to an uncompensated microcantilever beam by subjecting both the devices to a change in relative humidity (RH) at constant temperature and change in temperature at constant RH. This was done by using a commercial humidifier to introduce water vapor into the enclosing metallic box to achieve chamber humidification, a vacuum line for chamber evacuation, and a dry $N_2$ line for purging. The relative humidity of the chamber was monitored using a commercial chip, Bosch BME-280 connected to an Arduino Uno board. Temperature testing was performed by attaching a flexible heater to the chuck of the probe station and controlled using feedback received by a temperature controller from a thermocouple (TC) also attached to the chuck. Fig. 7a shows the schematic of the test setup used for this experiment.

*5.1.2 VOC testing*

In order to test the device against different VOC concentrations, the sensor was placed at the probe station, and a polycarbonate petri dish was used to cover the device. The VOC was delivered to this chamber using a Teflon tube. A small hole was cut out on the top of this petri dish to allow the micromanipulator probes to contact the chip, which also served as an exit path for the incoming gas. The required VOC concentration was obtained by bubbling dry $N_2$ through a conical flask containing the solvent, which was fed into a mixing chamber. The VOC vapors were diluted using a separate dry $N_2$ line, which was also attached to the mixing chamber. The diluted VOC concentration was then delivered to the device testing chamber using a Teflon tube at a constant flow rate of 2 liters/min. Fig. 7b shows a schematic of the gas delivery arrangement. The concentration of the VOC delivered to the device under test is given by,

$$C = \frac{\frac{P \times L}{P_o - P}}{\frac{P \times L}{P_o - P} + L + L_o} \times 10^6 \tag{4}$$

Where $P$ is the vapor pressure of the solvent (in mmHg) at room temperature, $P_o$ is the atmospheric pressure (in mmHg), $L$ is the flow rate of dry $N_2$ through the conical flask containing the solvent, $L_o$ is the flow rate of diluting dry $N_2$ into the mixing chamber and $C$ is the obtained concentration of the VOC in parts per million.

Self-leveling micromechanical VOC sensors with three different polymers (Polyimide, Polyurethane, and PDMS) were tested against five solvents; Acetone, Ethanol, Benzene, Hexane, and Water vapor using this arrangement.

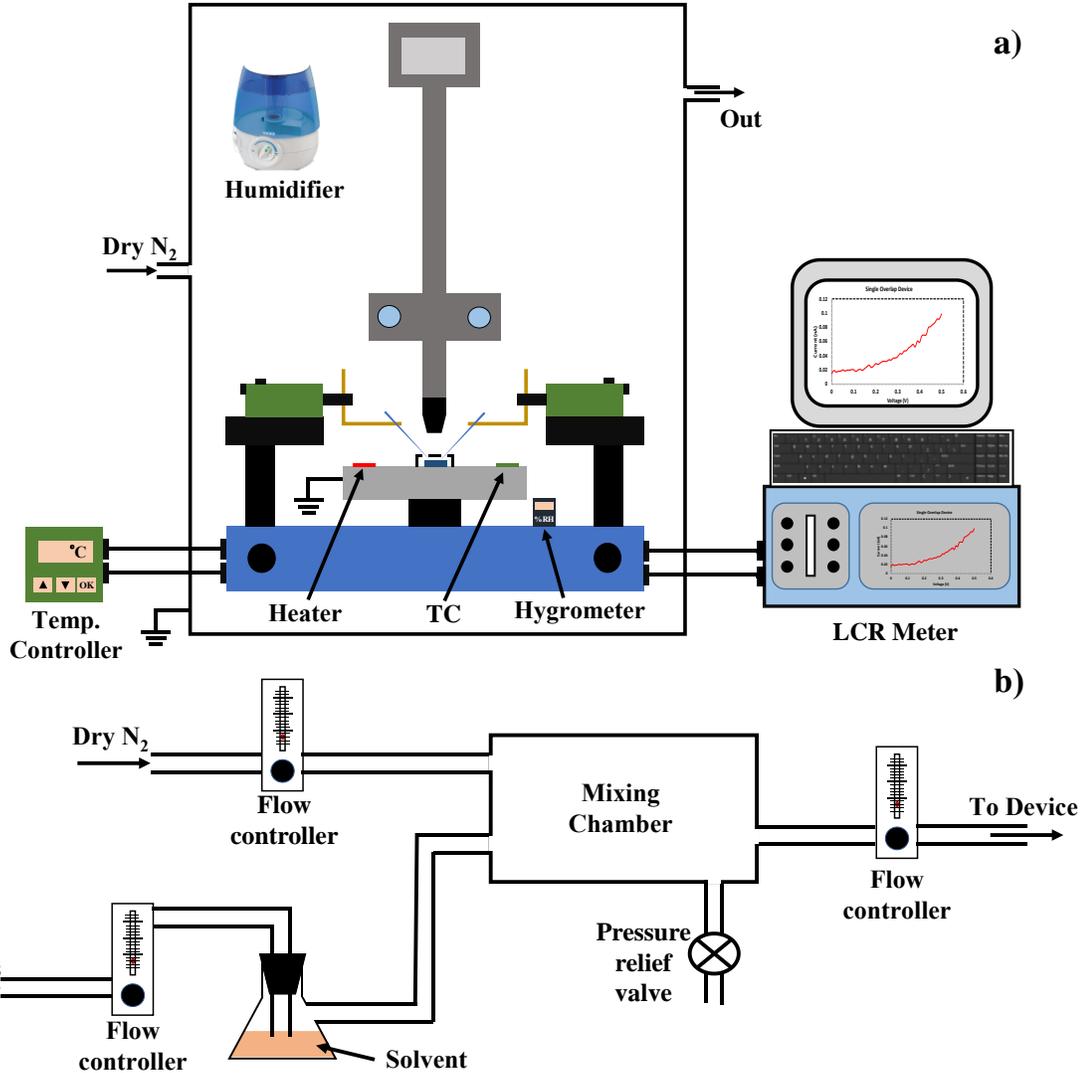

**Figure 7:** Schematic of the a) Device testing setup w.r.t relative humidity and temperature. VOC testing is done by delivering analyte vapor to a smaller chamber placed over the sample. b) Gas measurement setup for VOC sensing.

## 5.2 Sensor response to analyte and temperature

Fig. 8a shows the normalized response of the self-leveling Polyimide coated sensor as a function of varying water-vapor concentration compared to the response of an uncompensated reference microcantilever coated with Polyimide. The testing was repeated three times to plot the error bars from the calculated standard deviations. It can be seen that the self-leveling design has the same response to gas as that of a regular microcantilever sensor. However, the response of the self-leveling design to a temperature change is significantly different from a reference microcantilever, which shows about 47% drift when heated from 23-72°C. On the other hand, the self-leveling design shows a flat temperature response over the same temperature range, which is ~52x lower than the simple microcantilever as shown in Fig. 8b. The coefficient of hygroscopic swelling is approximately constant for T < 100°C [59]. Ideally, zero baseline drift can be achieved using the self-leveling geometry, but due to fabrication errors, this is not the case. After validating the gas and temperature response of the self-leveling design, three different polymers coated sensors were tested against different commonly available VOCs. Fig. 9a-b show the performance of the sensor when polyimide is used as a sensing polymer. It can be seen that the polyimide sensor shows high sensitivity to water vapor, medium response to ethanol and low sensitivity to Acetone, Benzene, and Hexane. Fig. 10a-b shows the response of the polyurethane-coated sensor. This device is much more sensitive to ethanol compared to the polyimide device and shows higher sensitivity to Benzene compared to Acetone. Fig. 11a-b shows the performance of the PDMS coated sensor to different analytes. This polymer is more sensitive to Acetone and Benzene compared to Hexane. PDMS is also more sensitive to VOCs overall compared to polyimide. It can be also seen that all the polymers show significant cross-sensitivity to different VOCs. However, this can be addressed using clustering algorithms as shown in the next section.

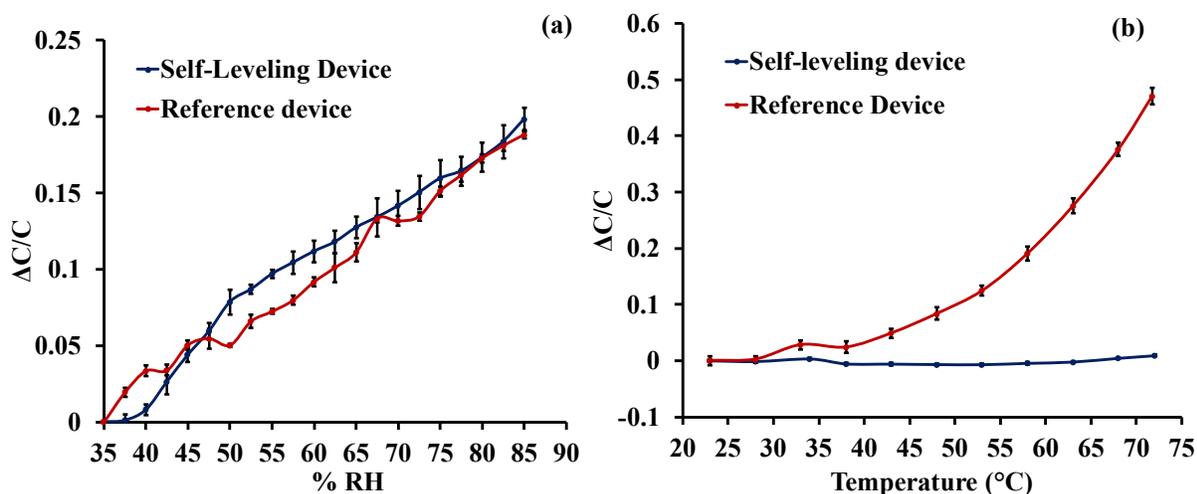

**Figure 8:** Comparison of polyimide-based temperature-compensated self-leveling sensor and uncompensated reference sensor a) Sensor response to varying %RH (at constant temperature ~23°C) b) Sensor response to varying ambient temperature (at constant ~ 42 %RH).

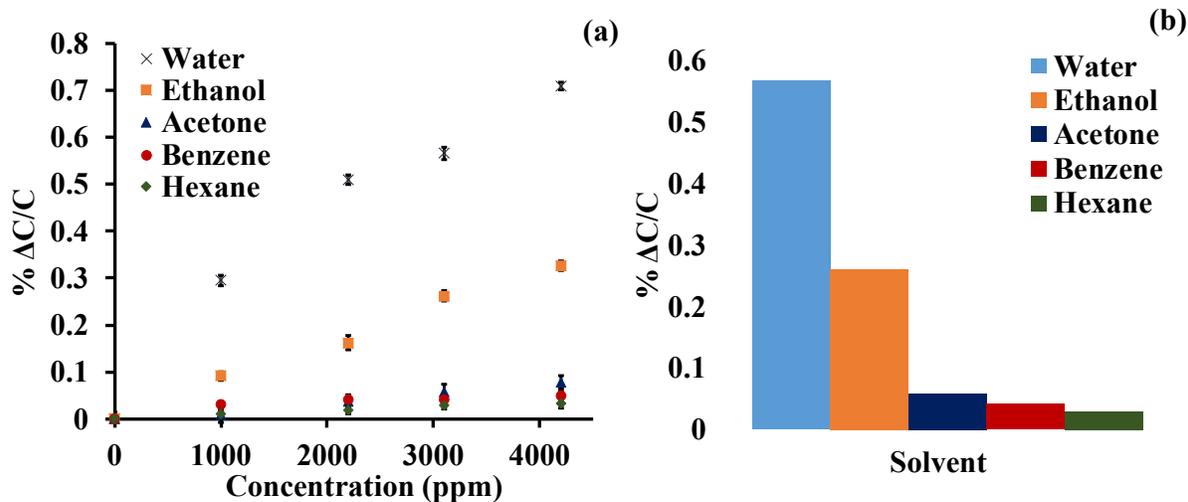

**Figure 9:** VOC response of the polyimide-based self-leveling sensor (a) Sensor response for varying VOC concentrations (b) Comparison of sensor response to different VOCs at 3100ppm concentration. Error bars show noise levels during measurement.

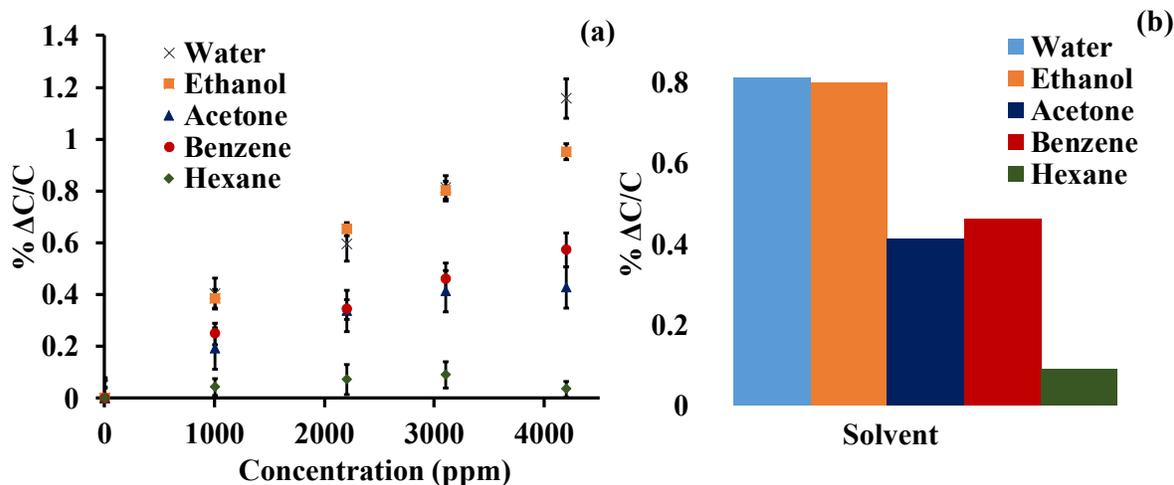

**Figure 10:** VOC response of the polyurethane-based self-leveling sensor (a) Sensor response for varying VOC concentrations (b) Comparison of sensor response to different VOCs at 3100ppm concentration. Error bars show noise levels during measurement.

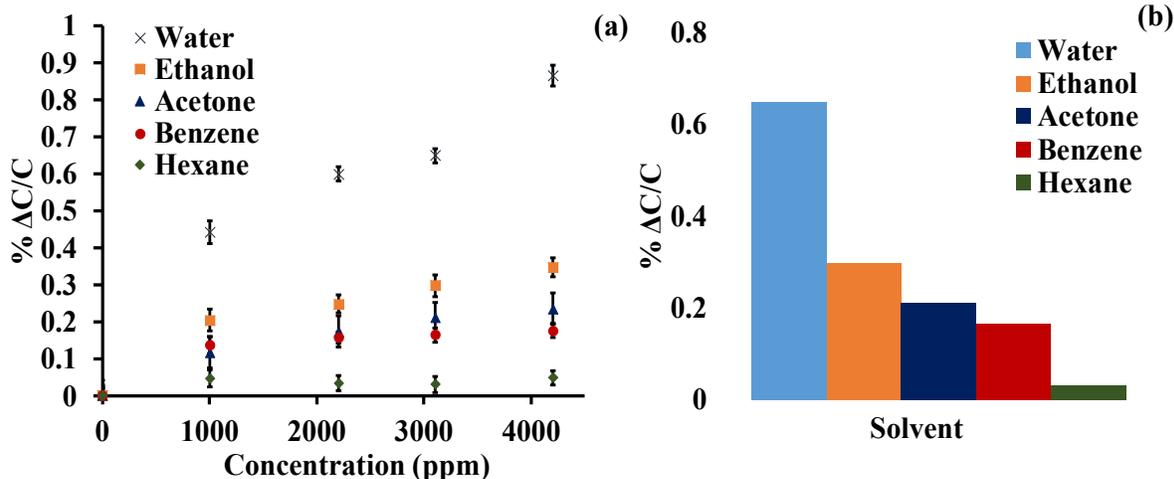

**Figure 11:** VOC response of the PDMS based self-leveling sensor (a) Sensor response for varying VOC concentrations (b) Comparison of sensor response to different VOCs at 3100ppm concentration. Error bars show noise levels during measurement.

*5.2.1 Classification Study*

The SVM study was implemented using the classification learner application in MATLAB 2019a. All the data points obtained during testing of the self-leveling VOC sensor for three polymers against five different analytes at 2200, 3100, and 4200ppm were used for this study. Since the dataset in this experiment was small, a 5-fold cross-validation scheme was used to prevent the overfitting of the generated model. Feature selection was done using PCA (Principal Component Analysis) to explain 95% variance of the data. It was observed that the Fine Gaussian SVM classifier gave the highest accuracy level of 95.3% for correctly identifying the analyte. Fig. 12 shows a confusion matrix obtained from this analysis, which compares the actual class of the analyte to the predicted class by the SVM based model for all the available data points. The percentage in Green and Red boxes indicate the success and failure rate of the model in classifying a particular analyte into its correct category. For example, out of the total data points, which were indeed Acetone, the model was able to correctly classify only 83.33% of the data points while 16.67% of the data points were incorrectly labeled as Benzene. The matrix clearly indicates that even though each of the single polymer-based self-leveling sensors show a significant degree of cross-sensitive behavior to different solvents as shown in Fig. 9b,10b, and 11b, use of appropriate pattern recognition algorithm allows selective detection of analyte based on the total response of multiple sensors.

*5.2.2 Response time*

Fig. 13 shows the response of the polyurethane-coated sensor to a step impulse of 3100ppm ethanol concentration. The response time $\tau_{90}$ for the device was defined as the time taken by the sensor to achieve 90% of its full-scale output response and was determined to be ~138s. The response time of the sensor is governed by diffusion of the analyte vapor into the polymer, which can be explained and modeled by a modified version of Fick's law of diffusion, as reported previously [22]. We believe the slow response the sensor is due to lack of adequate exposure sites of the polymer to the environment, as more than 70% of the surface area of the polymer on the inner beams is covered with the $Al_2O_3$ gas diffusion barrier. This can be improved by optimizing the trade-off between response time and temperature response of the device.

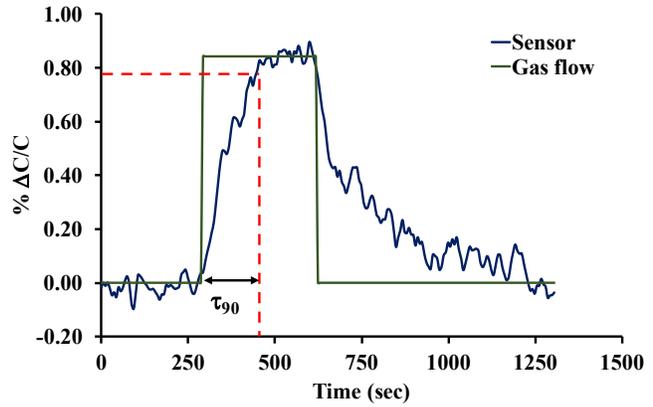

**Figure 12:** Confusion matrix obtained from a fine gaussian classifier based SVM clustering algorithm.

**Figure 13:** Step response of the Polyurethane coated self-leveling device to 3100ppm of ethanol.

## 6. Conclusion

We present the design, fabrication, and response of a batch-fabricated, capacitive, polymer-based VOC sensor based on mechanical leveraging. The device utilizes the self-leveling of MEMS microcantilevers to achieve passive temperature compensation to realize a polymer-based VOC sensor with near-zero baseline drift without the need for additional compensation sensors or electronics. This is the first report of a VOC sensor based on a self-leveling geometry. Self-leveling VOC sensors were microfabricated with three sensing polymers: Polyimide, Polyurethane, and PDMS to detect water vapor and different VOCs. We demonstrate that a self-leveling vapor sensor provides the same gas response as a simple microcantilever geometry, showing a ~20% change in device capacitance when subjected to 35-85 %RH change. Nearly-zero baseline drift due to changes in ambient temperature was demonstrated by the self-leveling VOC sensor when the temperature is increased from 23-72°C which is ~52-fold improvement over a conventional microcantilever geometry. This is a significant improvement over other vapor sensors with integrated temperature compensation reported in the literature [27,28,60] as the self-leveling vapor sensor does not require externally powered electronics to achieve temperature compensation. We also demonstrate the use of self-leveling vapor sensors for detection of different VOCs along with an SVM based classifier algorithm for selective detection of analytes with 95.3% accuracy. The VOC sensitivity of this vapor sensor can be improved further using an optimized sensor design and sense polymer to show better detection performance compared to state of the art low-power VOC sensors [61–63]. The dynamic response of the sensor was also characterized in this paper, and the sensor showed an absorption response time of ~138s. The results provided in this article show a clear justification to substitute simple microcantilever geometry based vapor sensors with self-leveling vapor sensors that are immune to temperature-induced baseline drift.

**Acknowledgements**

The authors would like to thank the staff at the University of Utah Nanofab for their assistance in fabrication of the devices. We would also like to thank Zhiheng Liu at the Department of Physics, University of Utah for assistance with SEM imaging of the devices.

**Funding**

This research did not receive any specific grant from funding agencies in the public, commercial, or not-for-profit sectors.

**Conflicts of Interest**

The authors declare no conflict of interest.